# Tracing Uncertainties of Extreme El Niño Prediction to Mesoscale Ocean Currents

Bo Qin[1], Zeyun Yang[2], Qiaofei Xue[3,4], Yanan Ding[2], Jinyu Wang[5], Chaopeng Ji[6], Shijin Yuan[1], Bin Mu[1], Wansuo Duan[3,4], Guokun Dai[5], Mu Mu[5], Yi Deng[7]

[1]School of Computer Science and Technology, Tongji University, Shanghai, China

[2]Key Laboratory of Ministry of Natural Resources for Marine Environmental Information Technology, National Marine Data and Information Service, Ministry of Natural Resources, Tianjin, China

[3]State Key Laboratory of Earth System Numerical Modeling and Application, Institute of Atmospheric Physics, Chinese Academy of Sciences, Beijing, China

[4]College of Marine Sciences, University of Chinese Academy of Sciences, Beijing, China

[5]Key Laboratory of Polar Atmosphere-ocean-ice System for Weather and Climate, Ministry of Education, Department of Atmospheric and Oceanic Sciences & Institute of Atmospheric Sciences, Fudan University, Shanghai, China

[6]CEREA, ENPC, EDF R&D, Institut Polytechnique de Paris, Paris, France

[7]School of Earth and Atmospheric Sciences, Georgia Institute of Technology, Atlanta, GA, USA

These authors contributed equally: Bo Qin and Zeyun Yang
*Corresponding author: Shijin Yuan (shijinyuan@tongji.edu.cn)

## Abstract

The predictability of the El Niño-Southern Oscillation (ENSO) has historically been linked to large-scale errors in models' initial conditions such as heat content and surface wind stress. Here we demonstrate that these slowly evolving errors may ultimately originate from mesoscale errors in initial oceanic states that are common in current models and ocean observation systems. This hidden source of ENSO forecast uncertainty, in terms of mesoscale ocean-current errors, is convincingly exposed by the conditional nonlinear optimal perturbation (CNOP) method applied to a differentiable ocean AI model. The CNOP-identified most influential current errors are strongly event-dependent. For the 1997/98 El Niño, the initial current errors in the eastern Pacific results in the largest forecast error through locally suppressing original subsurface warming. For the 2015/16 event, the dominate error originates from the western Pacific, which reorganizes upper ocean temperature structure and affects the eastern Pacific through the equatorial waveguide. Further analysis reveals that these current errors perturb mesoscale eddy count and eddy kinetic energy triggering large-scale temperature errors that hinder El Niño development. The timing of the maximum growth rate of the obtained CNOP error is well aligned with the spring predictability barrier of ENSO, and the error structures are reflected in simulated analysis errors caused by unresolved mesoscale currents. These findings extend the potential source of ENSO predictability beyond large-scale thermal signals, reveal for the first time an uncharted route of ENSO forecast failure, and provide an important physical basis for targeted ocean observations and data assimilation in the future.

## Introduction

The El Niño-Southern Oscillation (ENSO) is the dominant mode of interannual climate variability in the Earth system, shaping weather and climate extremes worldwide[1,2]. Yet ENSO is not a simple periodic phenomenon: individual event differs substantially in amplitude, spatial pattern, temporal evolution and teleconnections. Such diversities have become increasingly evident in recent decades, with central-Pacific, eastern-Pacific and mixed-type events posing severe challenges for both dynamical and statistical forecast systems[3]. To this end, further understanding ENSO predictability remains necessary in climate science[4]. One of the core tasks is to identify the initial errors that can evolve into large forecast uncertainties. Such dynamically sensitive errors provide valuable physical insights for recognizing forecast failures, guiding oceanic targeted observations[5], and enhancing data assimilations.

Previous studies have shown that initial errors with specific patterns strongly shape ENSO forecast uncertainties and are closely linked to the spring predictability barrier (SPB) in El Niño prediction[6,7]. These errors have been identified in a range of variables and models, including three-dimensional upper ocean temperature errors in the Community Earth System Model[8], joint sea surface temperature (SST) and sea level errors in an intermediate coupled model[9], and convection-sensitive signals that modulate the Walker circulation and Bjerknes feedback[10]. Despite the significant contribution of previous studies, these results mainly focused on monthly averaged and large-scale initial error. It is reasonable to suspect such initial error may be triggered by smaller-scale oceanic processes. Therefore, tracing the physical origins for such initial errors would require a detail research of mesoscale ocean dynamics.

Growing observation and simulation evidence suggests that smaller-scale ocean dynamics, particularly ocean current variabilities, can modulate ENSO evolution. Sub-mesoscale eddies near equator can inhibit the growth of both El Niño and La Niña events by influencing the upward heat flux from the subsurface to the surface ocean[11], while subsurface turbulent mixing can alter heat fluxes across the mixed layer and shape the monthly SST response of ENSO[12]. Numerical experiments further show that ocean resolution and mesoscale parameterizations can modify equatorial waves, background currents, eddy activity, ENSO asymmetry, and the amplitude and temporal variability of simulated events[13]. These findings suggest that monthly ENSO forecast errors may partly originate from mesoscale oceanic processes. However, it remains unclear whether specific structures of mesoscale ocean current errors can be responsible for the largest forecast uncertainties in ENSO events, and the relations between mesoscale oceanic processes and ENSO forecast accuracy remain largely unknown.

To reveal how mesoscale oceanic events such as the current initial error influences ENSO forecasts requires both a rigorous mathematical framework and a forecasting system capable of representing cross-scale oceanic variabilities. The conditional nonlinear optimal perturbation (CNOP) method[14] provides an approach combining mathematics and physics by formulating a nonlinear optimization problem which seeks a constrained initial perturbation resulting in the largest forecast error at a prescribed lead time. It has been widely used to address first-kind predictability problems in weather and climate systems[15]. The forecasting system, however, should simultaneously represent mesoscale ocean current variability and the interannual evolution of ENSO, which poses a great challenge. Conventional dynamical models used for ENSO simulation and prediction[16] typically have ocean resolutions of around 1°, and directly increasing their resolution would make CNOP-related optimization prohibitively expensive. Meanwhile, most existing data-driven ENSO forecast models only provide monthly averaged or large-scale multivariate predictions[17–22], limiting their ability to resolve mesoscale ocean processes and their links to ENSO variability. This leaves a methodological gap: the forecasting system used to optimize mesoscale initial errors should be both dynamically informative and computationally tractable.

Recent advances in ocean observations and reanalysis products create an opportunity to address this gap. Surface current fields can be quantitatively estimated from satellite-altimetry-derived sea-surface height products such as AVISO, and next-generation altimeters such as SWOT are providing unprecedented views of finer-scale ocean dynamics. At the same time, high-resolution ocean reanalysis products such as Global Ocean Reanalysis and Simulations (GLORYS) offer increasingly long, spatially refined and physically consistent records of ocean states. These datasets provide a large-sample foundation for training data-driven models[23] that can resolve both mesoscale ocean current variability and interannual ENSO evolution. When implemented as a differentiable forecasting system, such a model also enables efficient CNOP optimization through automatic differentiation[24]. This opens a practical pathway for identifying optimal mesoscale current error patterns and tracing their cross-scale influence on ENSO forecast uncertainties.

In this investigation, a differentiable, data-driven ocean forecasting model is developed and coupled with CNOP method. This framework is conducive to identify the optimal mesoscale ocean current errors that generate the largest ENSO forecast uncertainties, while retaining a physically interpretable connection between initial current errors and subsequent oceanic evolution. Applying this framework to 1997/98 and 2015/16 extreme El Niño events, the most influential current errors of these two events are strongly event-dependent. The 1997/98 event is most sensitive to current errors in the eastern Pacific ($5^{\circ}N-14^{\circ}N, 124^{\circ}W-96^{\circ}W$), which alter upper ocean temperature and local Bjerknes feedback, whereas the 2015/16 event is most sensitive to current errors in the western Pacific ($4^{\circ}N-10^{\circ}N, 136^{\circ}E-168^{\circ}E$), which reorganize upper ocean temperature and transmit their influence eastward through the equatorial waveguide. Further analyses show that these CNOP-type errors are associated with changes in mesoscale eddy activities within key regions, typically characterized by a continuous increase in cyclonic (cold) eddies during the first three months of error growth. Additional experiments further suggest that these errors are not simply products of the optimization procedure, but can constitute a potential source of the SPB, and are also reflected in simulated analysis errors arising from unresolved mesoscale currents. More broadly, these findings suggest that ENSO predictability cannot be fully understood from large-scale perturbations alone, but should also account for dynamically sensitive mesoscale current errors that connect unresolved ocean variability to climate-scale forecast uncertainty.

# Results

### Forecast skill of the differentiable data-driven ocean model

The constructed model covers the tropical-subtropical Pacific Ocean ($45^{\circ}S-45^{\circ}N, 112^{\circ}E-68^{\circ}W$) and predicts three upper ocean variables: temperature ($T$), zonal current ($U$) and meridional current ($V$). It is trained on subsets of the GLORYS high-resolution ocean reanalysis, with a horizontal resolution of $1/12°$ and 9 vertical levels in the upper 100 $m$ (0.49, 5.08, 9.57, 15.81, 25.21, 40.34, 55.76, 77.85 and 92.33 $m$). The original GLORYS data are averaged into pentad means to balance mesoscale ocean current variabilities and the slower interannual ENSO evolution, providing a suitable temporal representation for learning their cross-scale interactions. The training period spans 1993-2020, while the data from the subsequent 5 years are used for evaluation. To improve stability in long-term rollout forecasts, a two-stage training strategy is adopted.

Reliable scientific discovery with a data-driven model requires that the model retain sufficient forecast skill of the variables and timescales of interest. The basin-wide, multivariate forecast skill of the trained model is evaluated primarily. The model maintains effective forecast skill, defined by an anomaly correlation coefficient ($ACC$) greater than 0.5, for 87, 54 and 19 pentads for $T$, $U$ and $V$, respectively (Figure S1a-c). The corresponding root-mean-square errors ($RMSEs$) remain relatively low over these lead times (Figure S1d-f). A depth-wise evaluation further shows broadly consistent skill across all 9 vertical levels, with no pronounced degradation from surface to the subsurface ocean (Figure S1g-i). Additional regional and spatial evaluations, including assessments over the tropical, northern sub-tropical and southern sub-tropical Pacific and lead-specific skill spatial distributions, are provided in Figure S2 and Figure S3. These evaluations further show that the model retains coherent skill patterns, especially for $T$ and $U$, supporting its ability to represent both basin-scale variability and regional ocean-current structures.

The model also shows reliable skill in predicting ENSO evolution, as indicated in Figure 1. Its performance is compared with several operational forecast systems and an explainable data-driven model XOR. For the Niño 3.4 index, the constructed model maintains effective forecast skill for more than 1 year, with $ACC$ remaining above 0.5 for 81 pentads, suggesting that it captures the slowly evolving upper ocean memory that supports ENSO prediction. The forecast skill for Niño 4 is slightly higher than that for Niño 3, with both indices retaining skill beyond 1 year. This difference may be related to the stronger persistence of central-to-western Pacific SST anomalies and the greater influence of the SPB on eastern Pacific forecasts.

The forecast skills from different forecast start months are also evaluated (Figure 1d-e). The colored cells indicate lead times for which $ACC$ remains above 0.5, while the hatched cells mark the period of the fastest decline (i.e., top 20) in forecast skill within the 144-pentad forecast window. For all forecast starting calendar months, the model provides skillful forecasts for more than 8 months. Specifically, forecasts initialized during boreal autumn and winter generally show higher skill and often remain skillful for more than 1 year. The fastest decline in Niño 4 skill typically occurs during May to July, consistent with an early-summer predictability barrier for central Pacific forecasts[25]. In contrast, the fastest decline in Niño 3 skill occurs mainly during April to June, consistent with the classical SPB on eastern Pacific forecasts.

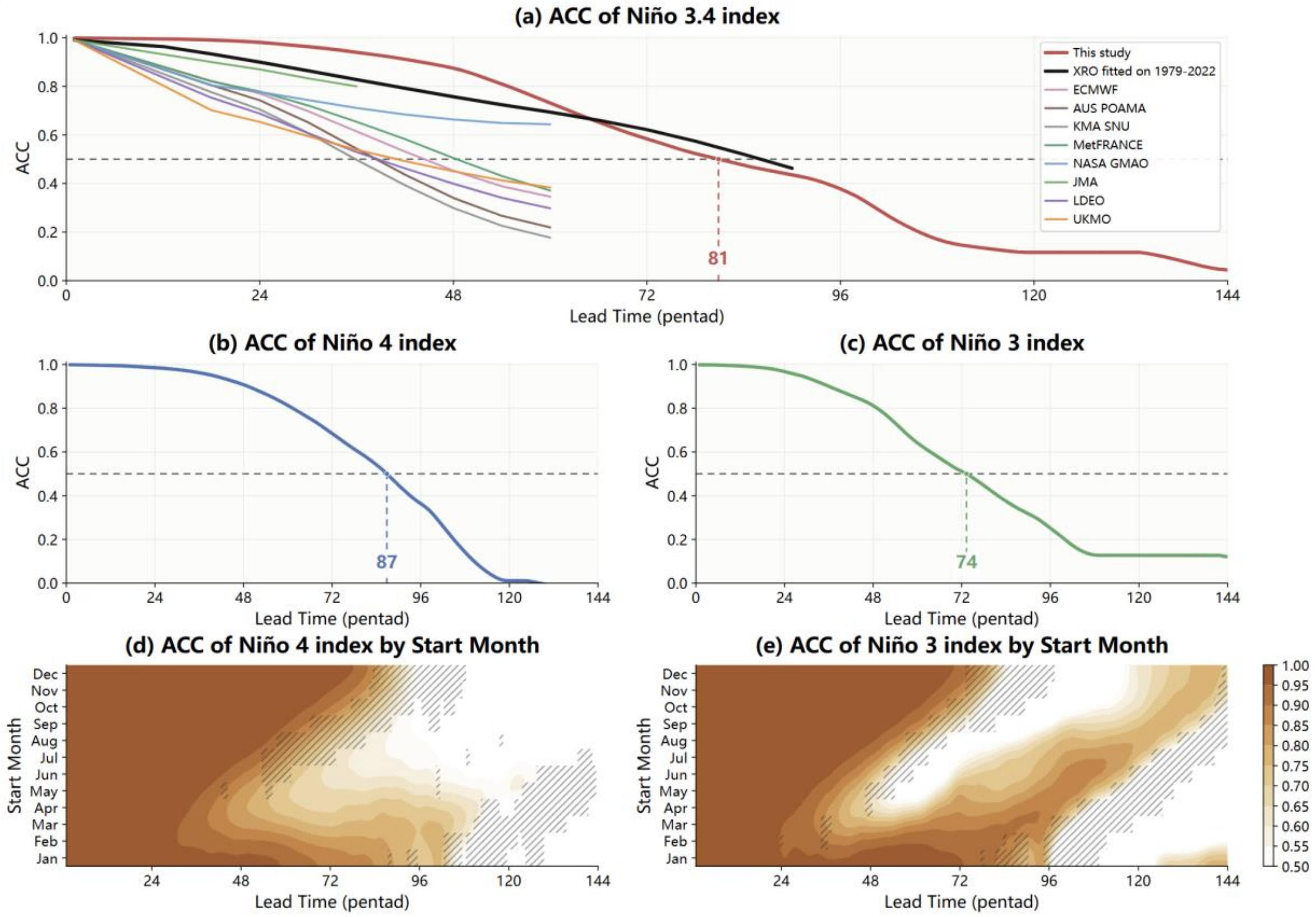

**Figure 1. ENSO forecast skill evaluated by Niño indices.** (a)-(c) show $ACC$ of Niño 3.4, Niño 4, and Niño 3 indices in different lead pentads, compared with operational forecast systems and the interpretable data-driven model XOR. (d) and (e) represent forecast skills from different forecast start months for Niño 4 and Niño 3, respectively. Colored cells indicate lead times for which $ACC$ remains above 0.5, and hatched cells mark the period of the fastest skill decline within the 144-pentad forecast window.

Overall, the model retains reliable skills over the equatorial and sub-tropical Pacific, where ENSO-related ocean variability is most active. $T$ remains skillful for about 1 year, $U$ for nearly 10 months and $V$ for approximately 4 months. These results, together with the coherent spatial skill distributions and effective Niño-index forecasts, indicate that the model can represent both mesoscale ocean current variability and interannual ENSO evolution. It can provide a robust forecasting and simulation system for diagnosing how mesoscale ocean current errors influence ENSO predictability.

## Optimal mesoscale ocean current errors for 1997/98 and 2015/16 El Niño events

The differentiable data-driven AI model is coupled with the CNOP method to identify the initial ocean current errors that can induce the largest ENSO forecast uncertainty. 1997/98 and 2015/16 extreme El Niño events are selected as two representative cases. Although they are included in the training period, the focus of this investigation is on ENSO predictability and error-growth mechanisms rather than independent forecast verification for individual events. Consistent with the seasonal phase locking of El Niño development, perturbations are superimposed on the ocean currents ($U$ and $V$) at all vertical levels during the developing phase, i.e., the last pentad of December 1996 and 2014 for the 1997/98 and 2015/16 cases, respectively. Starting from this perturbed initial state, the model is integrated forward for 72 pentads. The optimization objective is to maximize the forecast error (measured by $RMSE$) of SST over the Niño 3.4 region at the end of the integration, subject to an amplitude constraint on the initial perturbation. This constraint is limited to only 5% of the mean amplitude of ocean current anomalies in the training set, representing a small analysis-error magnitude and enabling an assessment of whether subtle current uncertainties can grow into large ENSO forecast errors.

For both 1997/98 and 2015/16 El Niño events, the monthly Niño 3.4 index evolution shows that the CTRL runs closely track the reanalysis (Figure 2a-b), confirming that the model can reproduce the development of these extreme events. However, when CNOP-type of initial current errors are superimposed, the predicted El Niño events are markedly weakened and can even result in neutral conditions. Thus, a small yet optimally structured perturbation in the initial ocean currents is sufficient to transform the predicted evolution of an extreme El Niño event, revealing mesoscale current errors as an efficient pathway for generating large ENSO forecast uncertainty.

The spatial patterns, i.e., vertically integrated oceanic kinetic energy of CNOP-type initial error, of the optimized perturbations further show that the most influential ocean current errors differ remarkably between the two events (Figure 2c-d). For the 1997/98 event, the vertically integrated oceanic kinetic energy is mainly concentrated in the eastern tropical Pacific and along the coastal region south of North America, shown in the black box in Figure 2c. In contrast, for the 2015/16 event, the vertically integrated oceanic kinetic energy is concentrated primarily in the western Pacific, especially near the western boundary of the North Equatorial Current and the western equatorial Pacific, shown in the black box in Figure 2d. This east-west contrast suggests that the optimal mesoscale ocean current errors are not characterized by a universal spatial pattern, but strongly related to the background state and growth pathway of each El Niño event.

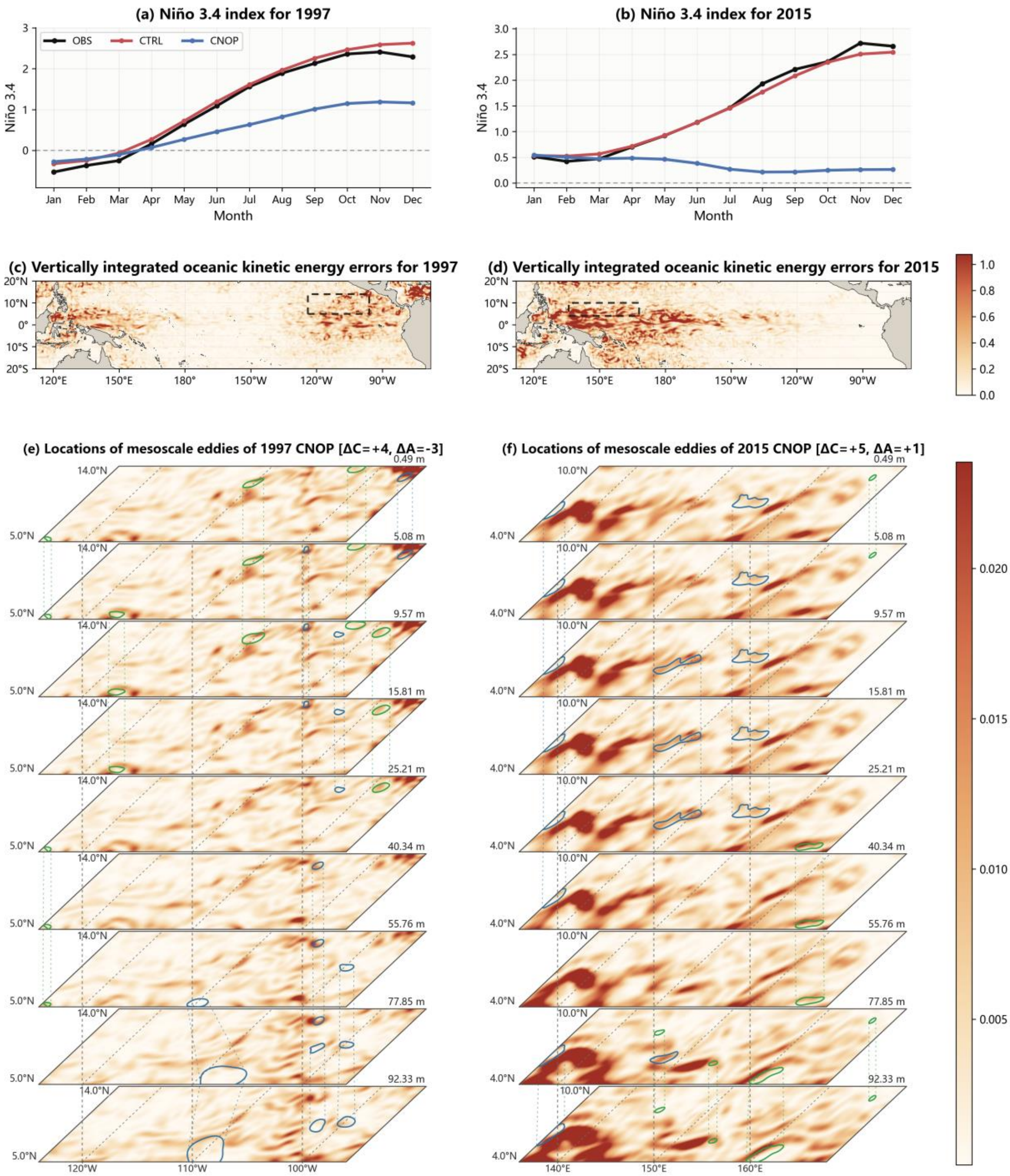

**Figure 2. CNOP-derived optimal ocean current errors for 1997/98 and 2015/16 El Niño events.** (a) and (b) represent monthly Niño 3.4 index evolution for the reanalysis (black line), the unperturbed control forecast (CTRL) and the forecast initialized with CNOP-type current errors (CNOP) for 1997/98 and 2015/16 El Niño events. (c) and (d) show the vertically integrated oceanic kinetic energy of the optimized initial ocean current errors for 1997/98 and 2015/16 events. Dashed boxes mark the key regions where the optimized current errors are concentrated and where subsequent mechanism analyses are performed. (e) and (f) describe three-dimensional mesoscale eddy structures identified within the key regions for 1997/98 and 2015/16 CNOP runs. Shading shows the depth-dependent oceanic kinetic energy, and colored contours mark detected cyclonic and anticyclonic eddies. The numbers in the panel titles indicate the net changes in cyclonic ($\Delta C$) and anticyclonic ($\Delta A$) eddy counts between the CNOP and CTRL runs.

Superimposing CNOP-type initial current errors produces distinct sea temperature anomalies evolutions in the two extreme El Niño cases. Figure S4 and Figure S5 show the CNOP-induced temperature errors, defined as the difference between the CNOP and CTRL runs, for the 1997/98 and 2015/16 events, respectively. For the 1997/98 event, the SST error response during the first three months is concentrated primarily in the eastern equatorial Pacific and the coastal region south of North America, where the error gradually intensifies from January to March. In the subsurface ocean, a negative temperature error appears from January below approximately 50 $m$ in the eastern equatorial Pacific and becomes increasingly organized by late March, indicating a localized growth of negative subsurface thermal error. In contrast, the western Pacific shows little coherent temperature error structure in either the surface or subsurface layers. After March, the negative temperature error continues to develop in the eastern Pacific through the following seasons, suggesting that the initial mesoscale current error can project onto a large-scale thermal response and progressively suppress the subsequent El Niño warming.

The 2015/16 case exhibits a rather different temperature-error evolution (Figure S5). During January and February, the CNOP-induced surface temperature errors are concentrated mainly in the western equatorial Pacific and near the western boundary region of the North Equatorial Current, while the eastern Pacific shows only weak and scattered responses. Meanwhile, the early subsurface errors are also rooted in the western Pacific. From March onward, these errors become increasingly organized and extend eastward along the equatorial Pacific. In the seasonal means, a pronounced west-east contrast emerges, with positive temperature errors in the western Pacific and negative errors across the central-to-eastern Pacific[26]. This evolution suggests that the 2015/16 current error pattern changes El Niño development through a tradition west-to-east error-growth pathway similar to previous studies, in contrast to the more locally amplified eastern Pacific response found in the 1997/98 case.

In general, these results indicate that CNOP-type ocean current errors can substantially reshape the predicted evolution of extreme El Niño events, but through distinct event-dependent pathways. The 1997/98 case is characterized by a locally amplified eastern Pacific response, whereas the 2015/16 case involves a western Pacific origin followed by eastward development. This contrast motivates a more detailed mechanism analysis of how mesoscale current perturbations are transformed into large-scale temperature and heat content errors through changes in oceanic kinetic energy and mesoscale eddy activity.

**Event-dependent growth pathways of optimal mesoscale ocean current errors**

Owing to the limitations in the variable completeness of the present data-driven model, employing energy-budget diagnostics to trace the underlying mechanisms is not feasible. Instead, this investigation applies two physically interpretable diagnostic quantities, eddy kinetic energy (EKE) error and upper ocean temperature error, to trace how mesoscale current perturbations are converted into large-scale thermal errors. A 120-day running mean is used to define the background flow[27], and vertical integration is employed to quantify the EKE error. Tests with other running-window lengths yield similar spatial patterns, indicating that the diagnosed error evolution is not sensitive to this choice. The upper ocean temperature[28] error is computed by vertically integrating the temperature difference between the CNOP and CTRL runs over 9 vertical layers, providing a measure of the accumulated thermal response to the optimized current perturbations.

The evolution of EKE errors and upper ocean temperature errors are shown in Figure 3 and Figure 4 for 1997/98 and 2015/16 cases, which further supports previous analysis of the two pathways for these two different cases. As shown in Figure 3, the EKE errors in the 1997/98 case are initially confined in the eastern tropical Pacific and the southern coastal region of North America, which is consistent with early development of the upper ocean temperature errors. This indicates that the shift of upper ocean temperature structure shown in Figure S4 is a result of the EKE errors. As the integration proceeds from January to March, the EKE errors intensify and expand over the eastern Pacific, while the associated upper ocean temperature errors continue to develop into pronounced negative structures over the eastern equatorial Pacific. By late March, the negative temperature errors become well organized then trigger local Bjerknes feedback to suppress the development of the original El Niño, resulting in a much weaker El Niño event compared with the CTRL run.

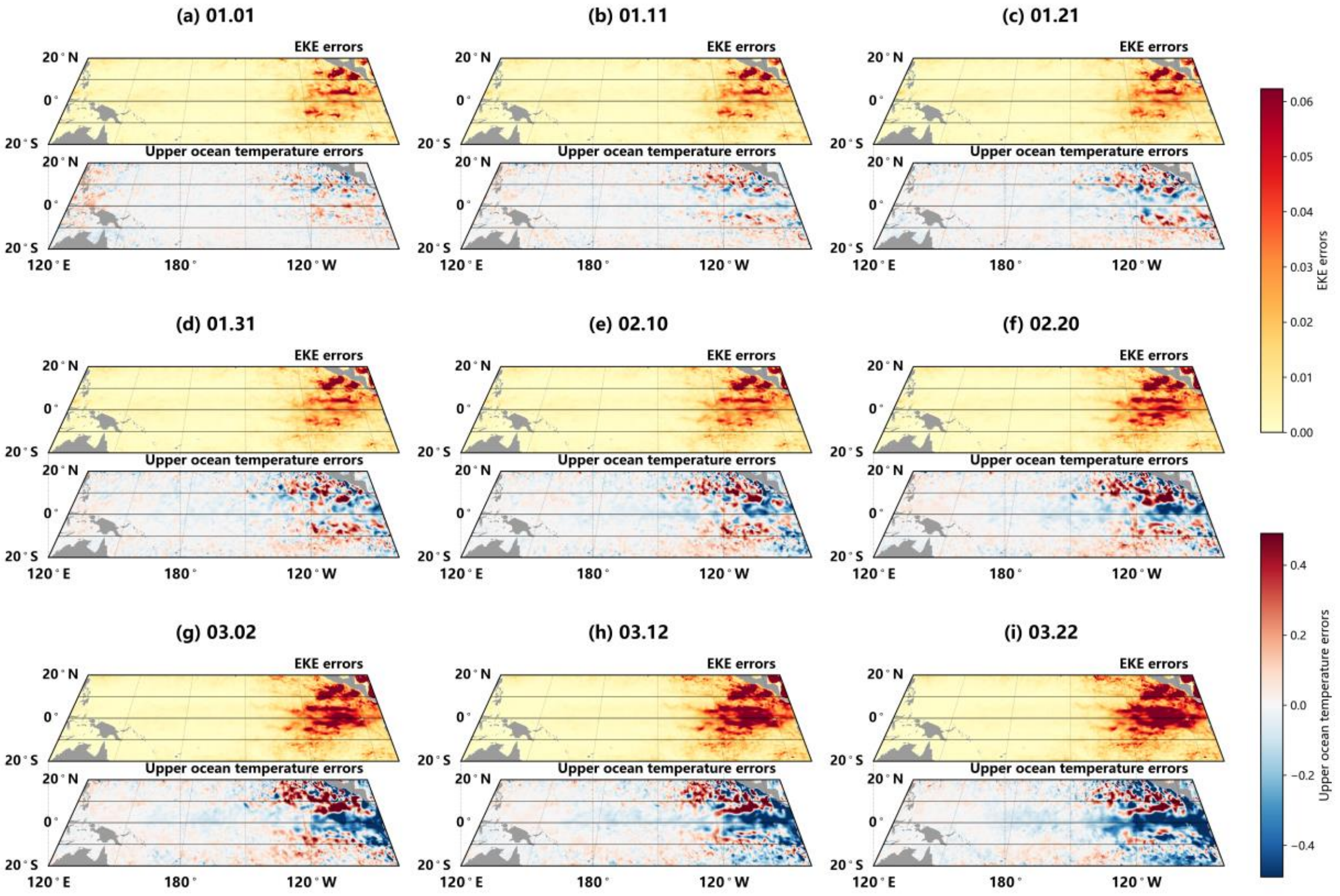

**Figure 3. Evolution of EKE errors and upper ocean temperature errors for the 1997/98 El Niño event.** (a)-(i) represent EKE errors in the upper maps and the corresponding upper ocean temperature errors in the lower maps.

In contrast, the 2015/16 case exhibits a western-Pacific origin of error growth (Figure 4). The EKE errors emerge near the western equatorial Pacific and the western boundary region of the North Equatorial Current, and then rapidly strengthen from January to March. The upper ocean temperature errors show a corresponding evolution: early errors are relatively scattered in January, but become increasingly organized over the western and central equatorial Pacific as the EKE errors intensify. By March, a clear west-east contrast emerges, with negative upper ocean temperature errors extending from the western-central Pacific towards the central basin and the positive errors located in the far western Pacific. This pattern is consistent with the equatorial averaged temperature error evolution shown in Figure S5, where the CNOP-induced response first appears in the western Pacific and subsequently develops eastward. These results indicate that, unlike the localized eastern-Pacific pathway in the 1997/98 case, the 2015/16 optimal current error influences ENSO prediction through a west-to-east pathway, in which mesoscale current perturbations are associated with reorganized upper ocean temperature structure in the western Pacific and the subsequent eastward development of thermal errors along the equatorial waveguide.

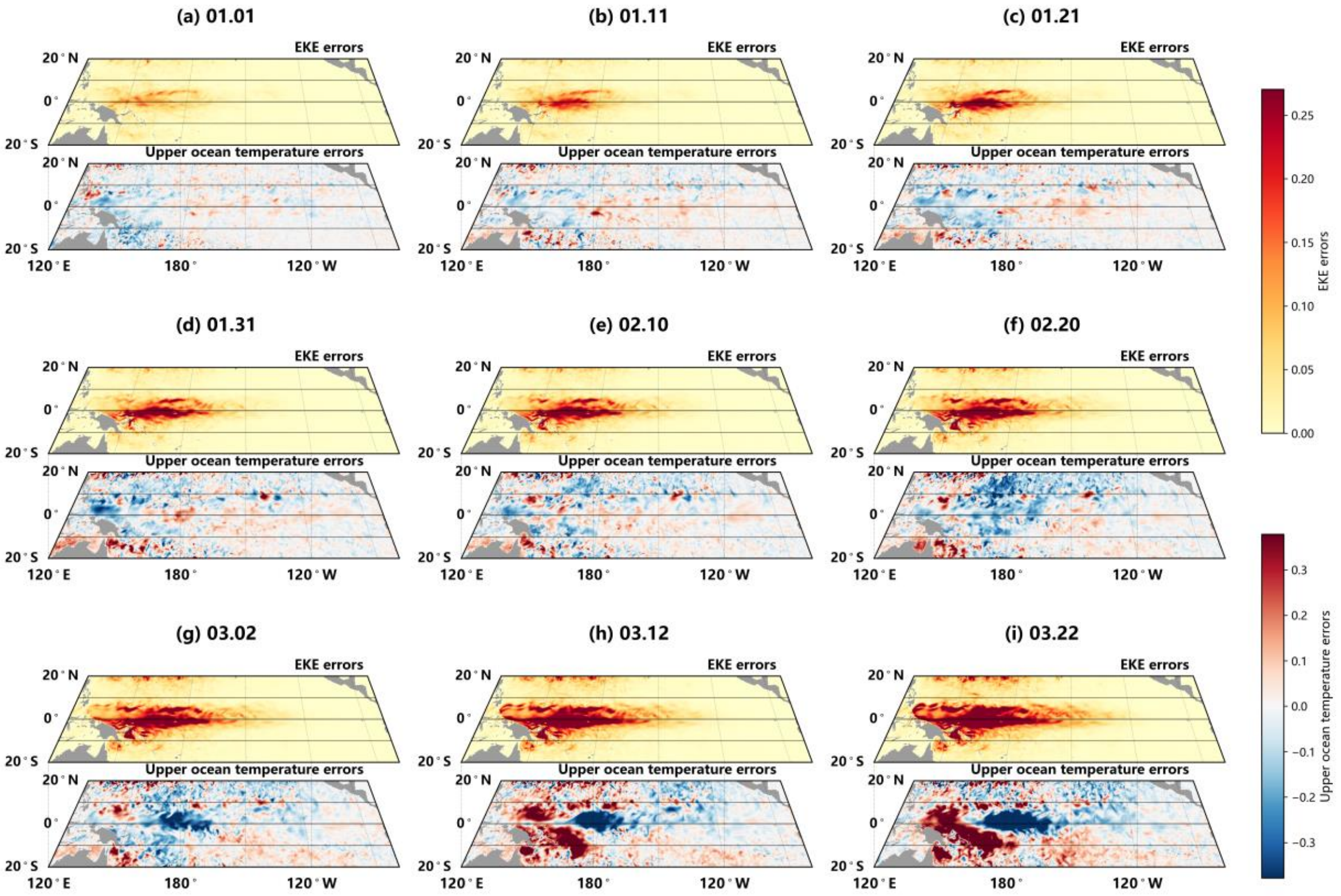

**Figure 4. Evolution of EKE errors and upper ocean temperature errors for the 2015/16 El Niño event.** (a)-(i) represent EKE errors in the upper maps and the corresponding upper ocean temperature errors in the lower maps.

The EKE and upper ocean temperature errors indicate that the optimized ocean current perturbations are associated with coherent thermal responses. However, EKE is a positive-definite quantity and cannot determine whether the associated mesoscale activity tends to cool or warm the upper ocean. To further connect EKE error with the sign of the upper ocean temperature response, three-dimensional mesoscale eddies within the key error-growth regions are identified. These regions are defined as areas in which EKE and upper ocean temperature errors show coherent amplification during the first three months. This procedure identifies an East region ($5^{\circ}N-14^{\circ}N, 124^{\circ}W-96^{\circ}W$) and a West region ($4^{\circ}N-10^{\circ}N, 136^{\circ}E-168^{\circ}E$), marked by dashed boxes in Figure 2b-c. The 1997/98 event is most sensitive to current errors in the East region, whereas the 2015/16 event is most sensitive to current errors in the West region.

Eddy identification is performed in these two key regions using the vector geometry (VG) method[29,30] based on the three-dimensional current fields. According to the properties of mesoscale eddies, anticyclonic eddies manifest as warm-core eddies, whereas cyclonic eddies manifest as cold-core eddies[31,32]. According to the identification result, the optimized initial current errors substantially modify the local eddy count (Figure 2e-f). In the 1997/98 case, the CNOP perturbation increases the number of cyclonic (cold) eddies by 4 and decreases the number of anticyclonic (warm) eddies by 3 in the East region, indicating a shift towards a locally cooling mesoscale eddy configuration. In the 2015/16 case, the CNOP perturbation produces an even stronger increase in cyclonic (cold) eddies in the West region, with 5 additional cyclonic (cold) eddies and only 1 additional anticyclonic (warm) eddy. These findings further confirm previous analyses that the optimized current errors can alter the thermal structure of upper ocean through vertical mixing of different types of mesoscale eddies.

To further verify the duration of these initial perturbations, the eddy count changes during the first three months are examined, which reveals a clear regional dependence of the CNOP-induced mesoscale response (Figure 5). For the 1997/98 case, eddy count changes in the West region are weak and irregular, consistent with the limited EKE and upper ocean temperature responses there. In contrast, the East region exhibits a more systematic adjustment during the first three months, with cyclonic (cold) eddies generally increasing and anticyclonic (warm) eddies generally decreasing. This persistent shift in eddy count is consistent with the development of negative upper ocean temperature errors in the eastern Pacific, suggesting that the optimized current error favors a mesoscale configuration that promotes upper ocean cooling processes. For the 2015/16 case, the most coherent eddy-count response shifts to the West region, where cyclonic (cold) eddies increase persistently, while anticyclonic (warm) eddies tend to decrease. This regional contrast matches the western-Pacific origin of EKE-error growth and the subsequent reorganization of upper ocean temperature errors. By comparison, the East-region response in 2015/16 is weaker and less persistent. In summary, these results reveal a cascade-like error-growth pathway from mesoscale flow perturbations to ENSO-scale thermal uncertainty. In this pathway, optimized current errors in event-dependent sensitive regions first alter mesoscale eddy activity and associated EKE structures, then reorganize regional upper ocean temperature structure, and ultimately generate large-scale temperature anomalies that suppress the predicted development of El Niño events.

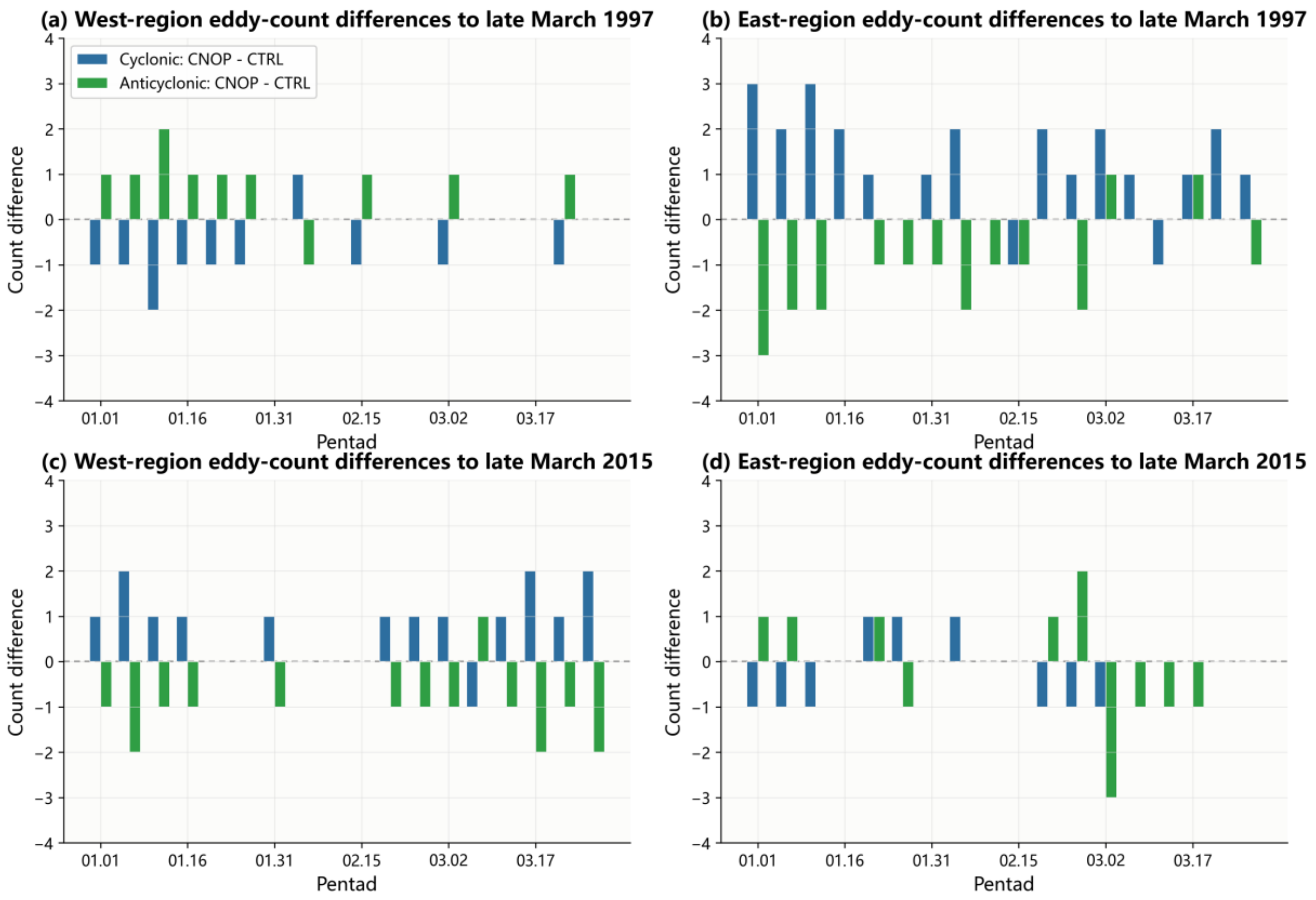

**Figure 5. CNOP-induced changes in mesoscale eddy counts within key error-growth regions.** (a) and (b) show eddy-count differences for the West and East regions in the 1997/98 case. (c) and (d) are the same as (a) and (b), but for the 2015/16 case.

## CNOP-type current error patterns in SPB and simulated analysis errors

Two optimal current error patterns are strongly event-dependent. For the 1997/98 El Niño event, which more closely resembles a canonical eastern-Pacific event, the optimized current error locates majorly at the East region. It directly modifies subsurface temperature in the eastern equatorial Pacific, thereby affecting the local Bjerknes feedback and suppressing the subsequent growth of eastern-Pacific warming. In contrast, for the 2015/16 event, which exhibits stronger central-to-western Pacific characteristics during its development, the optimized current error firstly reorganizes upper ocean temperature structure in the West region. Its influence then develops eastward along the equatorial waveguide and ultimately alters SST evolution in the eastern Pacific. These contrasting pathways indicate that mesoscale current errors do not limit ENSO prediction through a single universal pattern, but rather through event-dependent sensitive regions and growth routes.

Although these CNOP-type errors represent the most influential ocean current perturbations within the forecasting system, their broader significance depends on whether they contribute to understanding the limits of ENSO predictability and the reasonable initial errors. To this end, their growth is first examined to determine whether it exhibits the seasonal characteristics of the SPB[33–37], and similar structures are then investigated in simulated analysis errors associated with unresolved mesoscale currents.

To assess their SPB relevance, the monthly growths of objective function values are evaluated (Figure 6a-b). For the 1997/98 case, the error growth peaks sharply in April, consistent with the classical SPB for eastern-Pacific El Niño prediction. For the 2015/16 event, the strongest growth occurs from April to June, with substantial amplification persisting into early summer. This later and broader growth window is consistent with the stronger central-to-western Pacific influence of the 2015/16 event and with the west-to-east pathway diagnosed above. These results suggest that the optimized mesoscale current errors are not merely capable of generating large forecast errors, but are also dynamically connected to SPB of ENSO prediction.

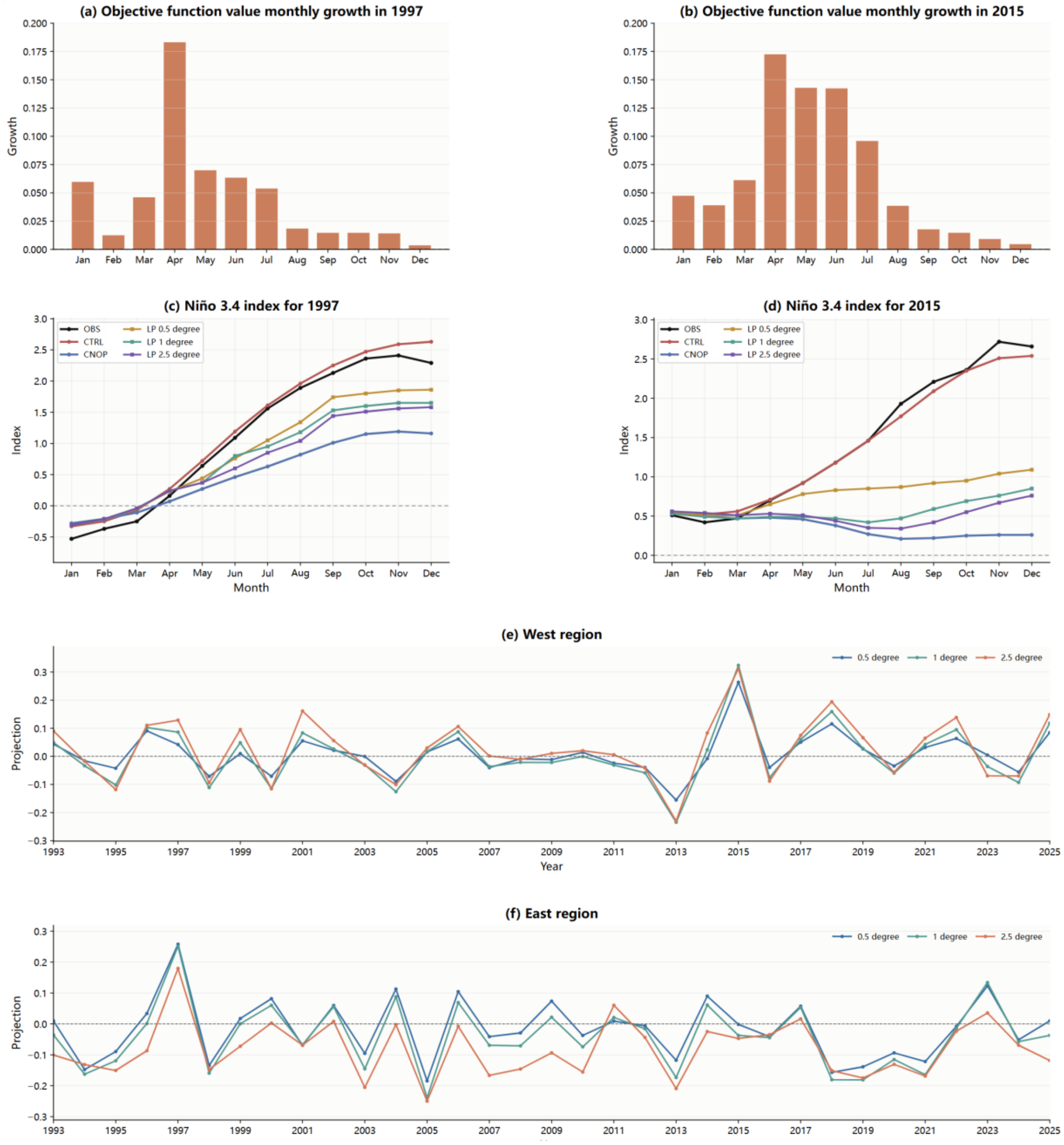

**Figure 6. SPB relevance and simulated-analysis-error signatures of CNOP-type current errors.** (a) and (b) represent the monthly growth of the Niño 3.4 SST forecast-error norm induced by CNOP-type initial errors for the 1997/98 and 2015/16 El Niño events. (c) and (d) show the monthly Niño 3.4 index evolution for the reanalysis, CTRL run, CNOP run and low-pass initialization experiments for the 1997/98 and 2015/16 events. The low-pass experiments use current fields smoothed at 0.5°, 1°and 2.5°scales to mimic the loss of unresolved mesoscale current structures. (e) and (f) describe the projection of simulated analysis errors onto CNOP-type current-error patterns in the West region and East region.

These optimized structures are further assessed by testing whether they are purely products of CNOP optimization or can be related to plausible analysis errors associated with unresolved mesoscale currents. Because true ocean analysis errors are not directly observable, simulated analysis errors are constructed by removing small-scale current structures from the high-resolution GLORYS data using low-pass smoothing at 0.5°, 1° and 2.5°. The difference between the smoothed and original current fields represents the current error that would arise if a forecast initialization system failed to resolve these mesoscale structures. Forecasts initialized with such smoothed current fields show weakened El Niño development compared with the CTRL runs, especially for the 2015/16 event, indicating that unresolved mesoscale currents can lead to non-negligible ENSO forecast errors (Figure 6c-d).

Meanwhile, the simulated analysis errors are projected onto the CNOP-type current-error patterns (Figure 6e-f). Although the absolute projection magnitudes are moderate, the projections show a robust event-dependent structure. In the East region, the 1997 case exhibits the strongest positive projection among all years and across the three smoothing scales. In the West region, the 2015 case shows the strongest positive projection. These results indicate that the CNOP-type current-error patterns are not simply mathematical optima constructed by the optimization procedure. Rather, components with similar structures can emerge from simulated analysis errors associated with unresolved mesoscale currents.

## Discussion

This investigation identifies the optimal mesoscale ocean current errors as a previously unexplored source of ENSO forecast uncertainty. By coupling a differentiable, data-driven ocean forecasting model with CNOP optimization, the optimal initial current errors that produce the largest forecast errors for 1997/98 and 2015/16 extreme El Niño events are identified. These errors are small in amplitude but dynamically efficient. They effectively hinder the predicted El Niño development and exhibit clear event dependence: the 1997/98 event is most sensitive to an eastern-Pacific current error pathway, whereas the 2015/16 event is most sensitive to a western-Pacific pathway that subsequently extends eastward along the equatorial waveguide. Further diagnosis reveals a cascade-like error growth pathway in which optimized current errors modify mesoscale eddy activity, structure, EKE, reorganize upper ocean temperature structure, and ultimately generate large-scale temperature errors that suppress El Niño warming. These results suggest that ENSO forecast uncertainty cannot be fully understood from large-scale thermal or thermocline errors alone and it also depends on dynamically sensitive mesoscale current errors embedded in the upper ocean.

Beyond this specific ENSO application, this investigation provides an example of how AI approaches can be used not only for operational forecasting, but also for mechanism demonstration in scientific discovery. The differentiable ocean AI model serves as a computationally efficient surrogate of high-resolution ocean model, while CNOP provides a physically interpretable optimization framework for extracting the most influential initial errors. This combination allows the model to move beyond black-box forecasting: it reveals sensitive structures, tests their forecast impact, and traces their evolution through physically meaningful diagnostics. The obtained current errors are not merely mathematical optima, as their growth has been shown to be connected to the SPB and their structures have been reflected in simulated analysis errors associated with unresolved mesoscale currents. In this sense, the AI model functions as a discovery tool that links high-dimensional oceanic data, nonlinear predictability theory and physically interpretable ENSO dynamics.

The results also have direct implications for targeted ocean observations and model initialization strategies. Previous targeted observation studies have emphasized that reducing dynamically sensitive initial errors can improve forecasts of high-impact weather and climate events. This investigation extends this idea to mesoscale ocean-current errors in ENSO prediction. The identified East and West regions represent event-dependent sensitive areas where small current errors can reorganize upper ocean temperature structure and affect subsequent SST evolution. Improving current observations and data assimilation in these regions may therefore help reduce ENSO forecast uncertainty, particularly when forecasts cross the seasonal window of the SPB. This issue may become increasingly important under climate change, as mesoscale eddy variability may alter in response to warming, stratification changes and circulation shifts[38–40]. In ENSO-sensitive regions, such changes could make upper ocean current errors more consequential for prediction. This investigation suggests that future ENSO observing systems should better constrain mesoscale current structures, alongside with traditional large-scale SST, thermocline and wind-stress signals.

Although the differentiable model resolves upper ocean temperature and currents at high resolution, it does not yet include the full set of oceanic and coupled air-sea processes. The related EKE, upper ocean temperature and eddy diagnostics provide an interpretable error-growth pathway rather than a closed momentum, energy or heat budget. The CNOP experiments also focus on two extreme El Niño events, so larger event ensembles are necessary to assess whether the identified East- and West-region sensitivities are general across different ENSO types. In addition, models that incorporate broader oceanic and air-sea processes will be required to further test these mechanisms and understand the potential atmospheric and oceanic sources of the mesoscale current errors. Such efforts would move the framework beyond the identification of sensitive errors towards resolving the full causal chain linking mesoscale current uncertainty, heat-content redistribution, air–sea feedback and ENSO predictability.

## Materials and Methods

### Data

The GLORYS high-resolution global ocean physical reanalysis provided by the Copernicus Marine Service are used as the primary data source. Three upper ocean variables within the tropical-subtropical Pacific Ocean ($45°S - 45°N, 112°E - 68°W$) are selected: sea temperature ($T$), zonal ocean current ($U$) and meridional ocean current ($V$). The horizontal resolution is $1/12°$, and the vertical levels used include 0.49, 5.08, 9.57, 15.81, 25.21, 40.34, 55.76, 77.85 and 92.33 $m$, representing the upper 100 $m$ of the ocean.

The data span 1 January 1993 to 31 October 2025. The original fields are averaged into pentad means to balance mesoscale ocean-current variability with the slower interannual evolution of ENSO, providing a temporal representation suitable for learning their cross-scale interactions. The samples are standardized using statistics estimated from the training period, and are divided into three subsets: a training set spanning 1 January 1993 to 31 December 2019, an evaluation set covering the full year of 2020, and an independent testing set spanning 1 January 2021 to 31 October 2025. The final dataset contains approximately 11988 samples after pentad averaging and overlapping sampling. Missing values over land were masked as 0.

The two extreme El Niño events analysed in the CNOP experiments are the 1997/98 and 2015/16 events. For both events, the initial state for perturbation optimization was taken from the last pentad of the year preceding the event peak (26-31 December), consistent with the seasonal phase locking of El Niño development.

## Model

A differentiable data-driven ocean forecasting model is constructed to predict the evolution of upper ocean temperature and currents over the tropical-subtropical Pacific (Figure 7). The model follows an Encoder-Processor-Decoder architecture and is designed to learn both mesoscale ocean-current variability and the slower interannual evolution associated with ENSO. The input consists of 6 consecutive pentads of three oceanic variables at 9 vertical levels. Thus, each input sample has the form $x_{in} \in \mathbb{R}^{6\times27\times1080\times2160}$, where $27 = 3 \times 9$ denotes the combined variable-level channels, and $1080 \times 2160$ denotes the horizontal grid. The model predicts the next-pentad ocean state $x_{out} \in \mathbb{R}^{27\times1080\times2160}$.

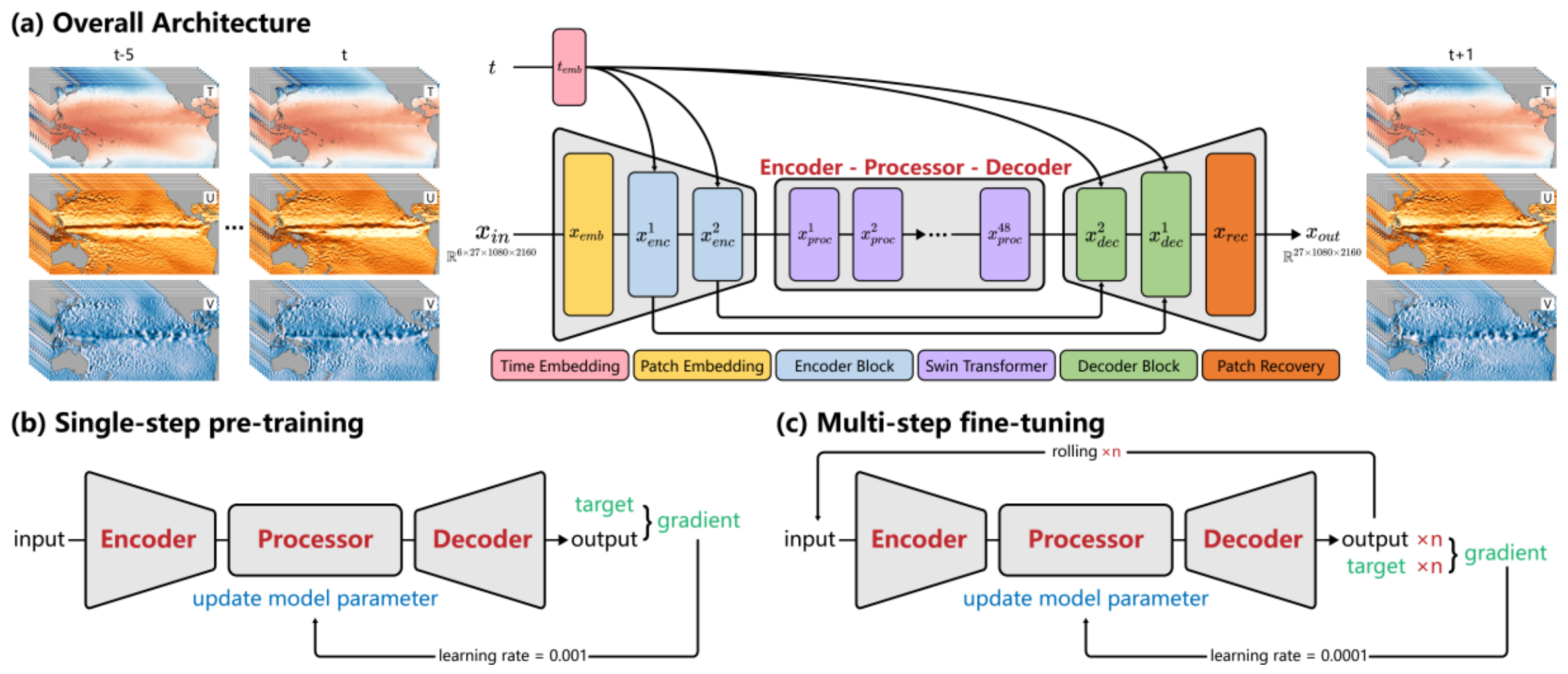

**Figure 7. Architecture and training strategy of the differentiable ocean AI model.** (a) is overall model architecture. (b) is single-step pre-training strategy. (c) is multi-step fine-tuning strategy.

As shown in Figure 7a, the input fields are first transformed into latent representations through patch embedding. A time embedding is incorporated to represent the temporal position of the input sequence and is injected into the encoder and decoder modules. The encoder consists of two encoder blocks that extract multiscale spatial features from the input ocean states. The processor is composed of 48 Swin-Transformer blocks[41], whose window-based self-attention allows the model to efficiently capture local mesoscale structures and broader spatial dependencies. The decoder then reconstructs the processed latent features through two decoder blocks, and a patch-recovery layer maps the latent representation back to the physical output fields. Module-level details of the patch embedding, patch recovery, encoder, decoder and Swin-Transformer blocks are provided in Figure S6.

The model is trained with a two-stage strategy using a mean-square-error loss to improve long-term roll-out stability (Figure 7b-c). In the first stage, single-step pre-training is performed, in which 6 input pentads are used to predict the next pentad. This stage allows the model to learn the basic spatiotemporal mapping between consecutive ocean states, with a learning rate of $10^{-3}$. In the second stage, multi-step fine-tuning is conducted by recursively feeding model outputs back as inputs for $n = 12$ forecast steps. The loss is accumulated over the rolled-out sequence, directly constraining multi-step error growth. The learning rate is reduced to $10^{-4}$ during this stage. Because all model components are differentiable, gradients of a scalar ENSO forecast-error objective can be propagated back to the initial current fields, enabling efficient CNOP optimization without requiring an adjoint model.

**Conditional nonlinear optimal perturbation method**

The conditional nonlinear optimal perturbation (CNOP) method[14] is used to identify the optimal initial ocean current error that can induce the largest ENSO forecast uncertainty within a prescribed amplitude constraint. Let $M$ be the nonlinear forecast model, $x_0$ be the unperturbed initial condition and $p$ be the initial perturbation applied to the ocean current components. The CNOP is defined as the perturbation $p^*$ that maximizes a forecast-error objective $J$ at a given lead time $\tau$ subject to an $\ell$-norm constraint $\varepsilon$ ($\ell = 2$):

$$p^* = \max_{\|p\|_\ell \leq \varepsilon} J(M_\tau(x_0 + p), M_\tau(x_0))$$

In this investigation, $p$ is imposed on $U$ and $V$ at all 9 vertical levels in the last input pentad. The forecast model is then integrated forward for 72 pentads. The forecast-error objective $J$ is to maximize the SST forecast error, measured by $RMSE$, over the Niño 3.4 region at the end of the integration. The obtained perturbation $p^*$ represents the initial current errors that most strongly affects ENSO development at a one-year lead time.

The constraint $\varepsilon$ is limited to 5% of the mean amplitude of current anomalies in the training set. This constraint represents a small analysis-error magnitude and is beneficial to examine whether subtle ocean current uncertainties can grow into large ENSO forecast errors. The optimization is solved using gradients obtained by automatic differentiation through the differentiable forecasting model.

**Anomaly correlation coefficient and root-mean-square error**

Forecast skill is evaluated using the anomaly correlation coefficient ($ACC$) and root-mean-square error ($RMSE$). Anomalies are computed relative to the pentad climatology $\bar{f}$ estimated from the training data. For a forecast result $\tilde{f}$ and the corresponding target $f$, $ACC$ is computed as

$$ACC = \frac{\sum_i (\tilde{f}_i - \bar{f}_i)(f_i - \bar{f}_i)}{\sqrt{\sum_i (\tilde{f}_i - \bar{f}_i)^2 (f_i - \bar{f}_i)^2}},$$

where $i$ denotes the grid points. $RMSE$ is computed as

$$RMSE = \sqrt{\frac{\sum_i (\tilde{f}_i - f_i)^2}{N}}.$$

where N denotes the grid count. A forecast is regarded as retaining effective skill when $ACC$ exceeded 0.5. Skill was evaluated at basin-averaged, depth-wise, regional and grid-point levels for $T$, $U$ and $V$.

**Niño indices and SPB-related error growth**

ENSO forecast skill is evaluated using Niño 3, Niño 4 and Niño 3.4 indices, computed as area-averaged SST anomalies over these three standard regions. Monthly index values are obtained from pentad forecasts by temporal averaging.

Besides, the seasonal growth of CNOP-induced ENSO forecast errors is examined by computing the monthly growth rate of the SST-difference norm between the CNOP and CTRL runs over the Niño3.4 region. The month of fastest error growth is then used to diagnose whether the optimized current errors exhibit characteristics of the SPB.

**Definition of key regions and projection of simulated analysis errors onto CNOP patterns**

Two key regions are defined based on the areas where vertically integrated EKE errors and upper ocean temperature errors show concurrent large-amplitude and coherent development during the first three months. The West region is defined as $4°N - 10°N, 136°E - 168°E$. The East region is defined as $5°N - 14°N, 124°W - 96°W$. These regions are used for eddy-count diagnostics and for projection analyses of simulated analysis errors.

To test whether CNOP-type current error patterns are reflected in plausible analysis errors, the simulated analysis errors are constructed by low-pass smoothing the high-resolution GLORYS current data at 0.5°, 1° and 2.5° scales and subtracting the original high-resolution data. These simulated errors represent the current errors that would arise if an initialization system failed to resolve mesoscale current structures. Their projection onto a CNOP $p^*$ is quantified by a normalized inner product:

$$P(e, p^*) = \frac{< e, p^* >}{\|e\|_2 \cdot \|p^*\|_2},$$

where $e$ denotes the simulated analysis error and the inner product $<\cdot,\cdot>$ is computed over the corresponding key region with vertical levels. Positive values indicate same-phase projection onto the CNOP-type initial error, whereas negative values indicate opposite-phase projection.

**Eddy kinetic energy and upper ocean temperature**

The eddy kinetic energy (EKE)-based metric is used to evaluate the error evolution[27]. The background flow $\bar{\cdot}_{120}$ is defined using a 120-day running mean. The mesoscale current anomalies are then computed as deviations from this background flow:

$$u' = u - \bar{u}_{120}, \quad v' = v - \bar{v}_{120}.$$

The EKE is calculated as

$$EKE = \frac{1}{2}(u'^2 + v'^2).$$

CNOP-induced EKE errors are computed as the difference between the EKE in the CNOP and CTRL runs. Sensitivity tests using alternative running-window lengths produce similar spatial patterns and does not affect the interpretation of the results.

The upper ocean temperature errors induced by the CNOP perturbation are computed from the vertically integrated temperature difference between the CNOP and CTRL runs over the upper ocean.

**Mesoscale eddy identification**

Three-dimensional mesoscale eddies are identified in the West and East regions using the vector geometry (VG) method[29,30] applied to the three-dimensional current fields. Eddy identification is performed separately at each vertical level, and the detected eddies are then aggregated over all vertical levels within each region. Cyclonic eddies are interpreted as cold eddies, whereas anticyclonic eddies are interpreted as warm eddies. Eddy-count differences are calculated as the total number of eddies in the CNOP run minus that in the CTRL run.

## Author contributions statement

B. Q., Z. Y., S. Y., B. M., W. D., M. M., and Y. D. provided the conceptualization and methodology. B. Q. and Q. X. contributed the model codes. B. Q. and Z. Y. conducted the experiments. B. Q., Z. Y., Q. X., Y. D. and G. D. conducted the validation, formal analysis, investigation, and visualization. S. Y., B. M., W. D., and M. M. provided the computation resource and funding. B. Q., Z. Y., and Y. D. wrote the draft. All the authors discussed the results, and reviewed and edited the manuscript.

## Competing interests

The authors declare no competing financial or non-financial interests.

Supporting Information for

# Tracing Uncertainties of Extreme El Niño Prediction to Mesoscale Ocean Currents

**Contents of this file**

Text S1
Figures S1 to S6

**Text S1. General evaluations of the differentiable ocean AI**

The basin-wide, multivariate forecast skill of the well-trained model is evaluated. The model maintains effective forecast skill, defined by an anomaly correlation coefficient ($ACC$) greater than 0.5, for 87, 54, and 19 pentads for $T$, $U$, and $V$, respectively (Figure S1a-c). The corresponding root-mean-square errors ($RMSEs$) remain relatively low over these lead times (Figure S1d-f), indicating that the model does not simply preserve correlation while accumulating large-amplitude errors. A depth-wise evaluation further shows broadly consistent skill across the 9 vertical levels, with no pronounced degradation from the near-surface layers to the upper subsurface ocean (Figure S1g-i).

Additional regional evaluations over the tropical, northern sub-tropical and southern sub-tropical Pacific are conducted (Figure S2). Overall, the forecast skill in the equatorial and northern subtropical regions is broadly consistent with the basin-averaged performance. In contrast, the southern subtropical region exhibits a slight degradation in forecast skill for all three variables. This reduction may be related to the more complex coastal geometry, island topography and energetic current systems in this region, which introduce stronger small-scale variability and make the local ocean state more difficult to predict.

The spatial distributions of forecast skill at lead times of 36, 48, 60 and 72 pentads are further examined (Figure S3a-l). For $T$ and $U$, forecast skill gradually decreases with lead time, but the regions of relatively high skill remain spatially coherent and largely unchanged. This persistence of high-skill regions indicates that the model does not merely reproduce basin-scale mean variations, but also retains meaningful predictive information for regional and multi-scale spatial features. In contrast, $V$ does not exhibit a clear and persistent skill pattern, suggesting weaker spatial organization or stronger internal variability. To interpret these differences, an index $I = {|C|}/{\delta}$ is defined, where $|C|$ represents the climatological magnitude of a variable and can be regarded as its dominant and deterministic structure, whereas $\delta$ denotes its variability and reflects the internally variable component that may hinder model learning. The vertical-averaged spatial patterns of this index for the three variables (Figure S3m-o) closely resemble their corresponding forecast-skill distributions at different lead times (Figure S3a-l). This consistency suggests that the predictability, or the upper limit of forecast skill, of each variable is strongly constrained by its intrinsic spatiotemporal characteristics, in addition to external factors such as model architecture and training strategy.

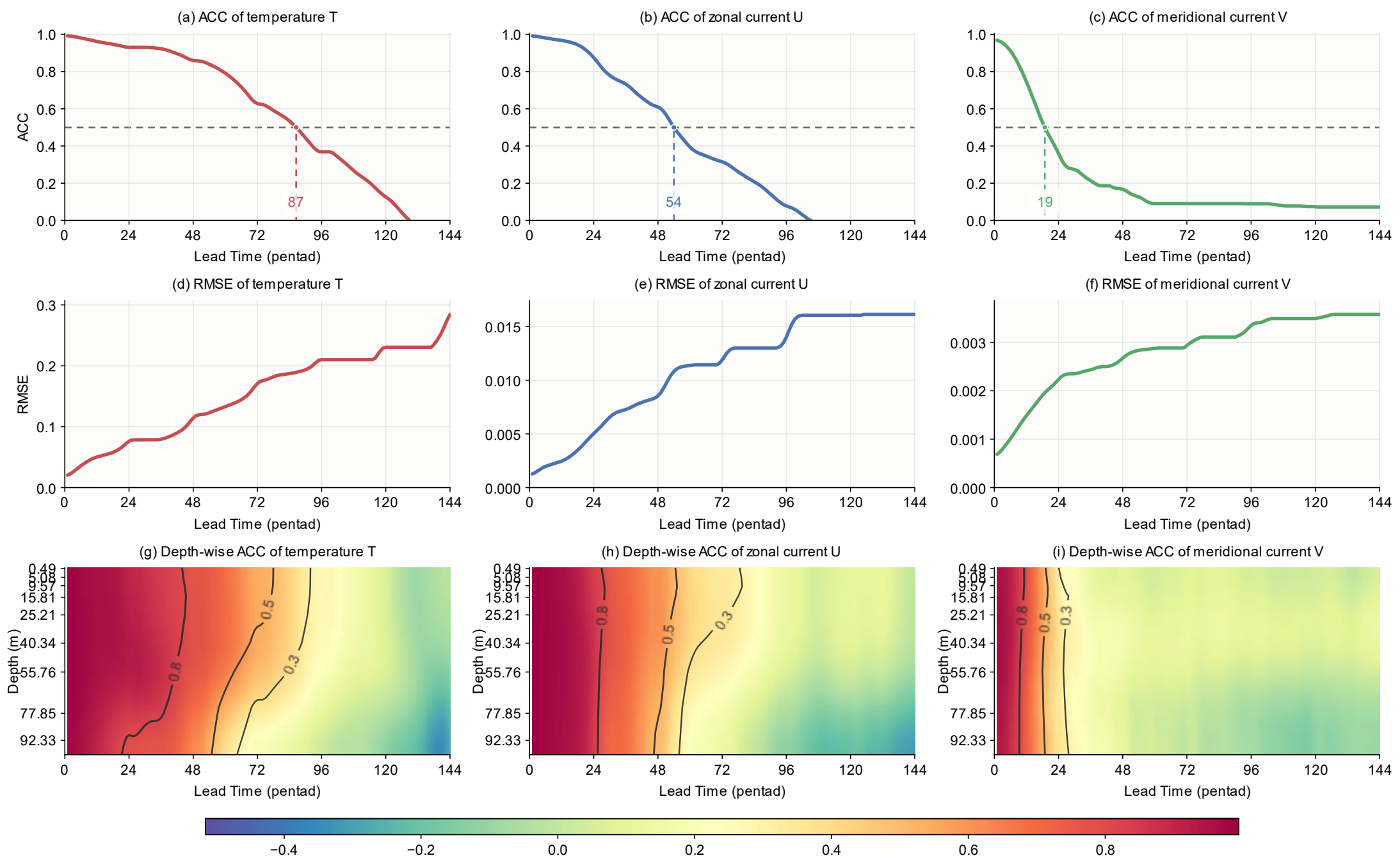

**Figure S1. Basin-averaged forecast skill of the differentiable ocean AI model.** (a)-(c) represent $ACC$ for $T$, $U$, and $V$. (d)-(f) represents corresponding $RMSE$. (g)-(I) shows depth-wise $ACC$ for $T$, $U$, and $V$ across the 9 vertical levels.

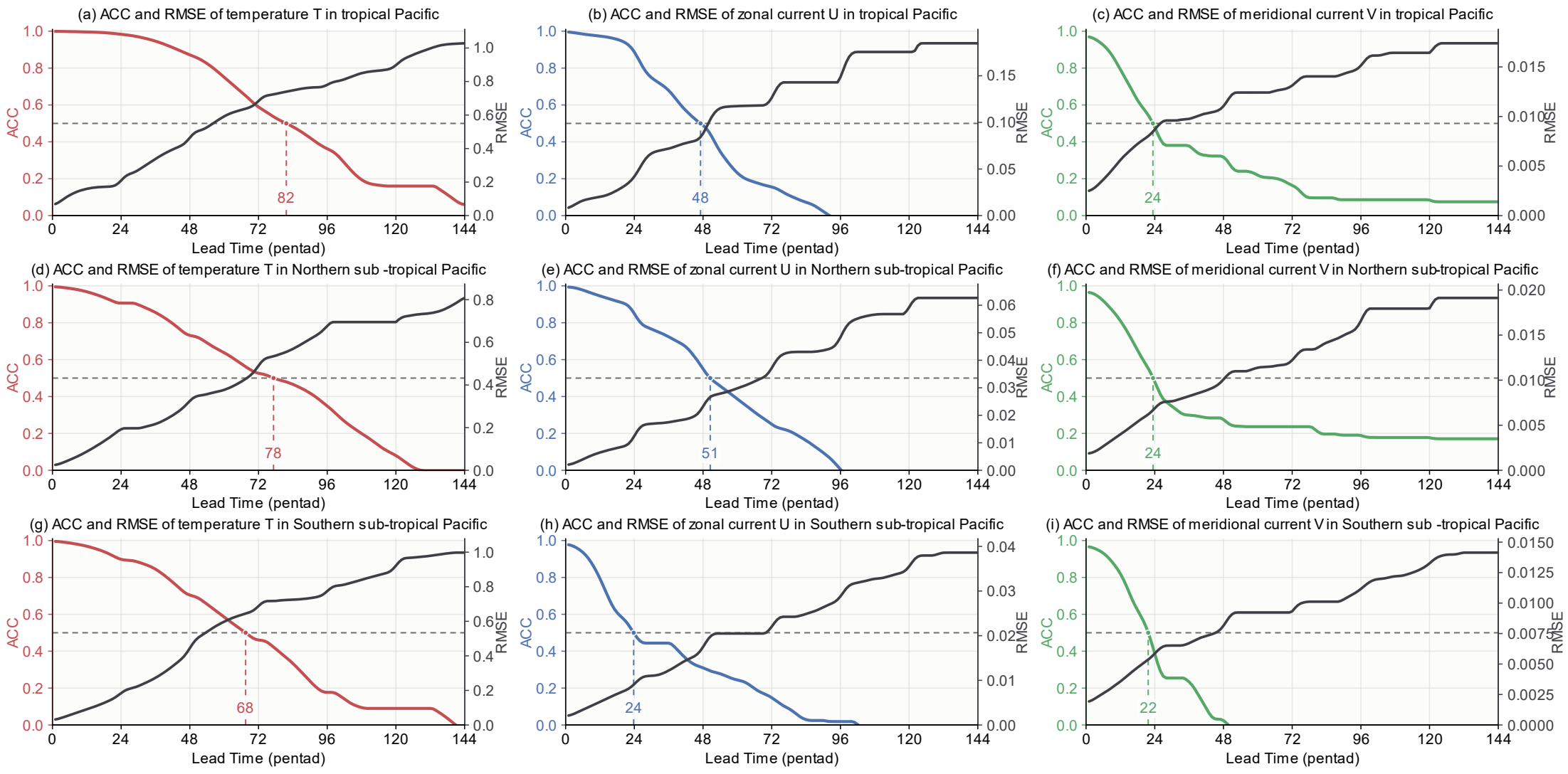

**Figure S2. Regional forecast skill over the tropical and subtropical Pacific.** (a)-(c) the tropical Pacific, (d)-(f) northern sub-tropical Pacific, and (g)-(i) southern sub-tropical Pacific. The $ACC$ and $RMSE$ are calculated for $T$, $U$, and $V$ in different lead pentads.

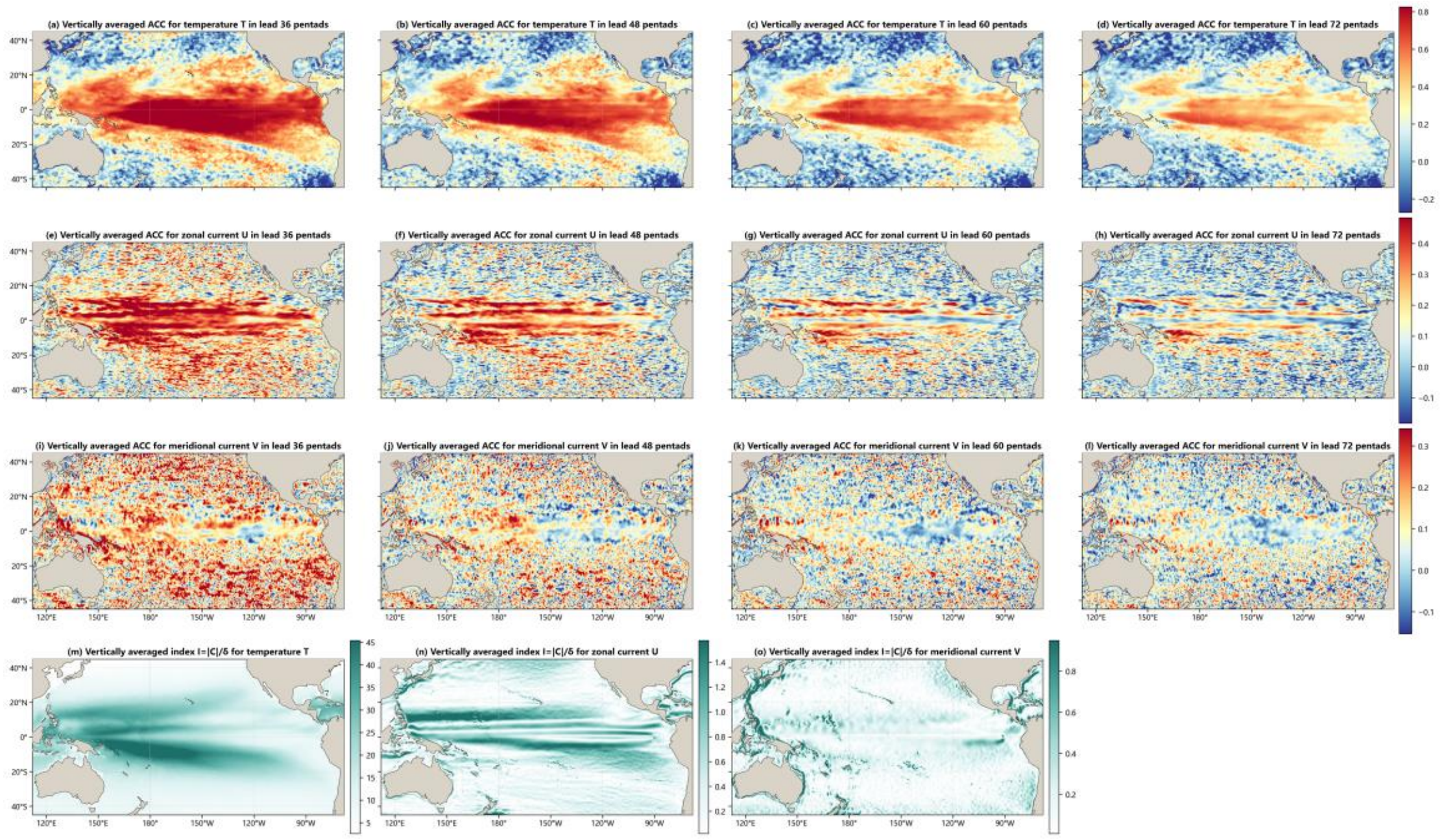

**Figure S3. Spatial distribution of forecast skill and intrinsic predictability index.** Vertically averaged spatial distribution of forecast skill for (a)-(d) $T$, (e)-(h) $U$, and (i)-(l) $V$ at lead times of 36, 48, 60 and 72 pentads. Forecast skill is measured by the $ACC$ at each grid point. (m)-(o) show the intrinsic predictability index $I = {|C|}/{\delta}$, where $|C|$ denotes the climatological magnitude of each variable and $\delta$ denotes its variability. Larger $I$ values indicate regions where the deterministic background structure is stronger relative to internal variability.

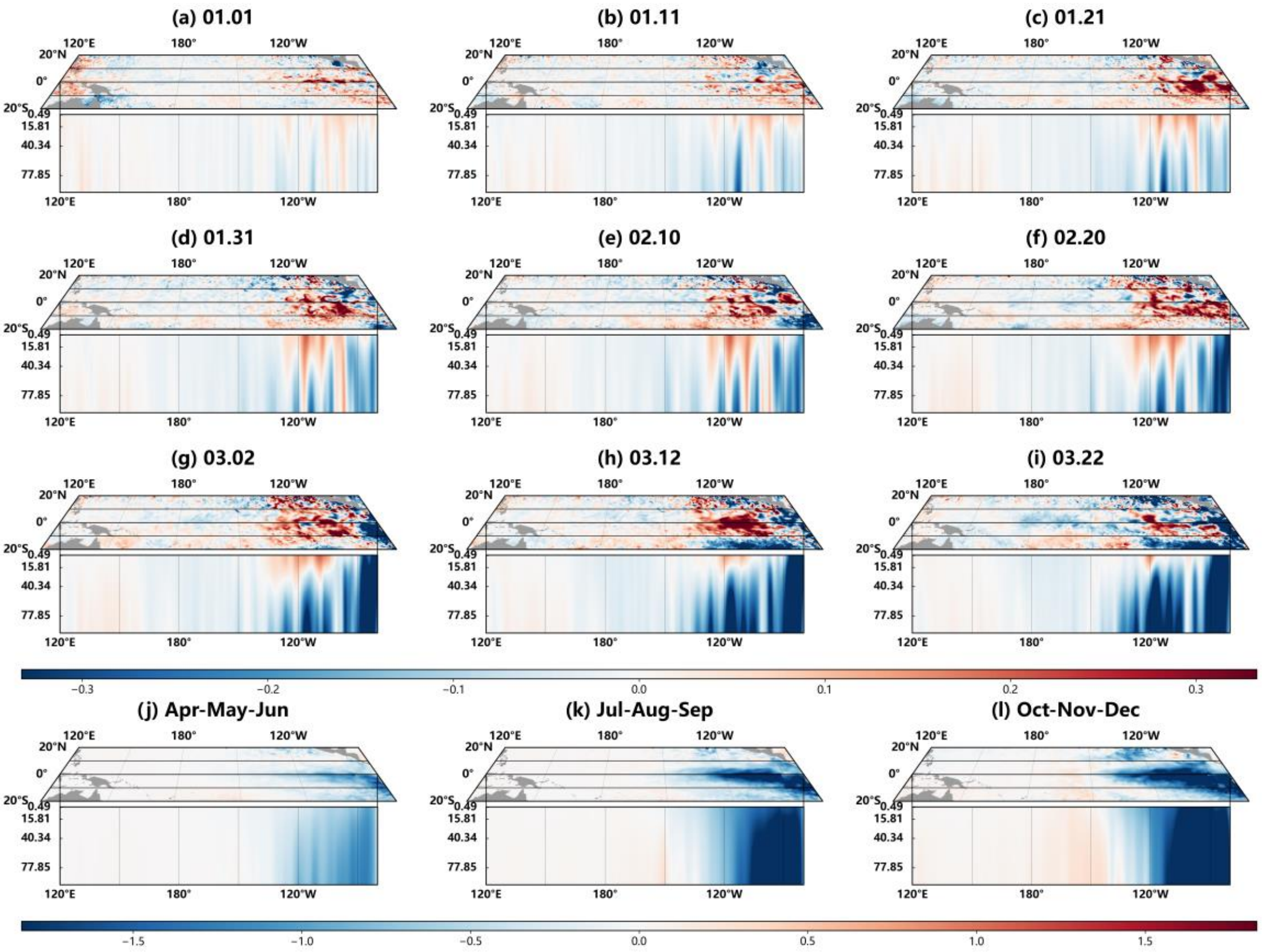

**Figure S4. CNOP-induced temperature-error evolution for the 1997/98 El Niño event.** Temperature errors are defined as the difference between the CNOP and CTRL runs. (a)-(i) represent evolution of temperature errors from 1 January to 22 March. Each panel shows the SST errors over the tropical Pacific in the upper map and the corresponding equatorial averaged temperature error over $5°S–5°N$ in the lower map. (j)-(l) show the seasonal-mean temperature errors for April-June, July-September and October-December, respectively. Shading denotes temperature errors in $°C$.

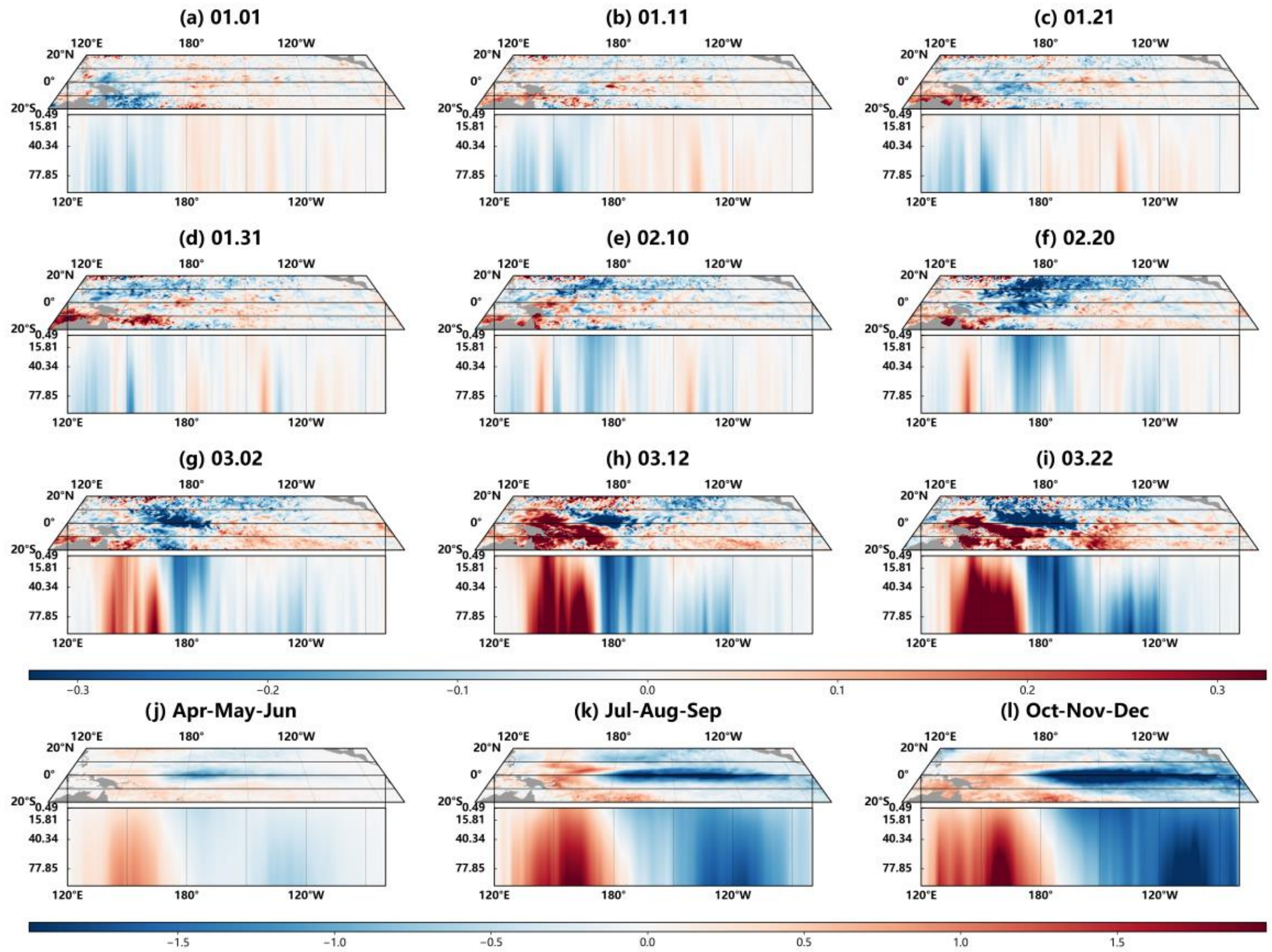

**Figure S5. CNOP-induced temperature-error evolution for the 2014/15 El Niño event.** Temperature errors are defined as the difference between the CNOP and CTRL runs. (a)-(i) represent evolution of temperature errors from 1 January to 22 March. Each panel shows the SST errors over the tropical Pacific in the upper map and the corresponding equatorial averaged temperature error over $5°S - 5°N$ in the lower map. (j)-(l) show the seasonal-mean temperature errors for April-June, July-September and October-December, respectively. Shading denotes temperature errors in $°C$.

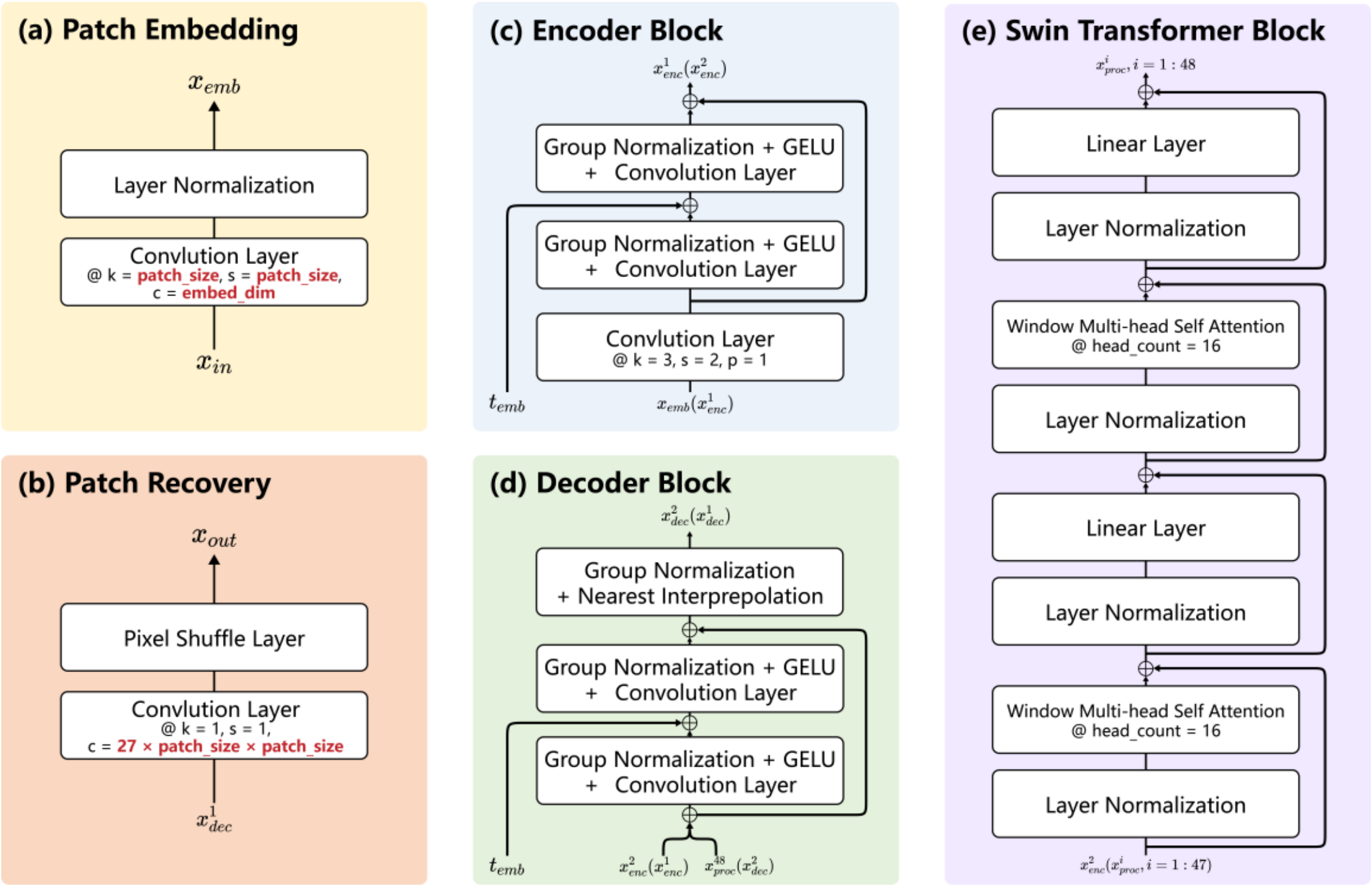

**Figure S6. Module-level architecture of the differentiable ocean AI model.** (a) is patch-embedding module. The input ocean fields are first projected into latent features using a convolutional layer with kernel size and stride equal to the patch size, followed by layer normalization. (b) is patch-recovery module. The latent decoder features are mapped back to the physical output fields using a convolutional layer and a pixel-shuffle layer. (c) is Encoder block. Each encoder block combines convolutional operations, group normalization, GELU activation and residual connections to extract multiscale spatial features from the embedded ocean states. Time embeddings are injected into the encoder block to represent the temporal position of the input sequence. (d) is decoder block. Each decoder block uses group normalization, GELU activation, convolutional operations, nearest-neighbor interpolation and residual connections to reconstruct spatial features from the processed latent representation. Time embeddings are also included in the decoder. (e) is Swin-Transformer processor block. The processor consists of repeated Swin-Transformer blocks with window-based multi-head self-attention, layer normalization, linear layers and residual connections. Each attention module uses 16 heads, and 48 processor blocks are used in the full model.